\documentclass[12pt]{article}
\usepackage{epsfig}
\def\be{\begin{equation}}
\def\ee{\end{equation}}
\def\bea{\begin{eqnarray}}
\def\ena{\end{eqnarray}}
\def\tg{\tilde{\gamma}}

\newcommand{\nn}{\nonumber\\}
\newcommand{\p}[1]{(\ref{#1})}

\renewcommand{\a}{\alpha}

\renewcommand{\c}{\gamma}

\renewcommand{\k}{\kappa}
\renewcommand{\l}{\lambda}
\begin{document}
\baselineskip=15pt
\begin{titlepage}
\setcounter{page}{0}
\begin{flushright}
KU-TP 020 \\
arXiv:0803.1013
\end{flushright}
%%
%% title
%%
\vspace*{5mm}
\begin{center}
{\Large \bf Cosmological Evolution of Dirac-Born-Infeld Field}\\
\vspace{15mm}
%%
%% author
%%
{\large Zong-Kuan Guo\footnote{e-mail address: guozk at phys.kindai.ac.jp}
and
 Nobuyoshi Ohta\footnote{e-mail address: ohtan at phys.kindai.ac.jp}}\\
\vspace{10mm}
 {\it
 Department of Physics, Kinki University, Higashi-Osaka,
 Osaka 577-8502, Japan}
\end{center}
\vspace{20mm}
%%
%% abstract
%%
\centerline{\large \bf Abstract}
{
We investigate the cosmological evolution of the system of a
Dirac-Born-Infeld field plus a perfect fluid.
We analyze the existence and stability of scaling solutions
for the AdS throat and the quadratic potential.
We find that the scaling solutions exist when the equation
of state of the perfect fluid is negative and in the
ultra-relativistic limit.
}
%%
%% PACS
%%
\vspace{2mm}
\begin{flushleft}
PACS number(s): 98.80.Es, 98.80.Cq
\end{flushleft}
\end{titlepage}
\setcounter{page}{2}

%%%%%%%%%%%%%%%%%%%%%%%%%%%%%%%%%%%
\section{Introduction}
\label{sec0}
%%%%%%%%%%%%%%%%%%%%%%%%%%%%%%%%%%%

Inflation in the early universe provides a natural explanation
for the homogeneity and isotropy of the universe and for
the observed spectra of density perturbations.
Recently inflationary models from string theory have
attracted much attention.
One approach to string inflation is based on D-brane~\cite{Quevedo}.
Of particular interest are scenarios where a type IIB
orientifold is compactified on a Calabi-Yau three-fold,
where the moduli fields are stabilized due to the
presence of non-trivial flux. These fluxes generate local
regions within the Calabi-Yau space with a warped
geometry or ``throat". In many settings, an anti-D3-brane
is fixed at one location in the infrared tip of
the throat and a mobile D3-brane experiences a small
attractive force towards the anti-D3-brane. The distance
between the branes plays the role of the inflaton field
and, since this is an open string mode, its dynamics is
determined by a Dirac-Born-Infeld (DBI) action.
Such a DBI action with higher derivative terms gives a
variety of novel cosmological consequences~\cite{sil03,che05,mar08,ben03}.

It is well known that, in a universe containing a perfect
fluid and a normal scalar field with an exponential potential,
for a wide range of parameters the scalar field mimics the
perfect fluid with the same equation of state~\cite{cop97}.
The scaling solutions in which the ratio of the energy densities of
the two components is a constant are realized in such a system and
are attractors at late times.
In tachyon cosmology, the inverse square potential for a tachyon
field allows similar scaling solutions, just like the exponential
potential does for a normal scalar field~\cite{pad02}.
This kind of scaling solutions are useful for explaining the current
acceleration of the universe.
It is thus interesting to investigate whether scaling solutions are also
present and stable in the DBI scenario.

In this paper, we undertake the first attempt to study a
system of dimensionless dynamical variables of the DBI field
plus a perfect fluid by using the phase-plane analysis method
which has been widely applied~\cite{iva03,tsu04,zha06}.
In the case of the AdS throat and the quadratic
potential, the system can be cast into an autonomous system.
We find that in addition to the DBI inflationary solutions, there
exist scaling solutions in the ultra-relativistic case.
We analyze their existence and stability.

%%%%%%%%%%%%%%%%%%%%%%%%%%%%%%%%%%%
\section{Autonomous System}
\label{sec1}
%%%%%%%%%%%%%%%%%%%%%%%%%%%%%%%%%%%

Consider the following effective action~\cite{sil03}
\bea
\label{action}
S = -\int {\rm d}^4x \frac{1}{g_{\rm YM}^2}
\sqrt{-g} \left[f(\phi)^{-1}\sqrt{1+f(\phi)
 g^{\mu\nu}\partial_\mu\phi\partial_\nu\phi} - f(\phi)^{-1}
 + V(\phi)\right]+S_{m}\,,
\ena
where $g_{\rm YM}^2$ is the Yang-Mills coupling
and $V(\phi)$ is a potential of the DBI field $\phi$.
In the case of the AdS throat, we have
$f(\phi)=\lambda/\phi^4$, where $\lambda$ is
the {}'t Hooft coupling which is related
to $g_{\rm YM}^2$ via the relation
$\lambda=g_{\rm YM}^2 N$ in the large-$N$ limit
of the field theory.
In the action (\ref{action}), we have also taken into
account the contribution of a perfect fluid.

In a spatially-flat Friedmann-Robertson-Walker (FRW) metric, the
energy density and pressure of the DBI field are given by
\bea
&& \rho_{\phi}=\frac{\c-1}{f}+V(\phi)\,,\\
&& P_{\phi}=\frac{\c-1}{f\c}-V(\phi)\,,
\ena
where
\bea
\c \equiv \frac{1}{\sqrt{1-f(\phi)\dot{\phi}^2}}\,.
\ena
The field equations read
\bea
\label{frieq}
&& H^2=\frac{\k^2}{3}\left[\frac{\c-1}{f}+V(\phi)+\rho_m\right], \\
\label{frieq2}
&& \dot{H}=-\frac{\kappa^2}{2} \left[ \gamma \dot{\phi}^2
+(1+w_m)\rho_m \right]\,,\\
&& \ddot{\phi} + \frac{3f_{,\,\phi}}{2f}\dot{\phi}^2
- \frac{f_{,\,\phi}}{f^2}
 + \frac{3H}{\c^2}\dot{\phi}
 + \left(V_{,\,\phi}+\frac{f_{,\,\phi}}{f^2}\right)\frac{1}{\c^3}=0\,,\\
&& \dot{\rho}_m+3H(1+w_m)\rho_m=0\,,
\label{evoeq}
\ena
where a dot denotes a derivative with respect to $t$ and
$\k^2=1/(g_{\rm YM}^2 M_p^2)$ with $M_p$ being the reduced Planck
mass. Note that $\rho_m$ and $P_m$ are the energy density and the
pressure of the fluid with an equation of state $w_m=P_m/\rho_m$.

We define the following variables:
\bea
& &
x \equiv \frac{\k}{\sqrt{3}H}\sqrt{\frac{\c}{f}} \,,\quad
y \equiv \frac{\k \dot{\phi} \sqrt{\gamma}}{H}\,, \quad
z \equiv \frac{\k \sqrt{V}}{\sqrt{3}H}\,, \nonumber \\
& &
\mu_1(\phi) \equiv \frac{V_{,\phi}}{\kappa f^{1/2}V^{3/2} }\,,
\quad
\mu_2(\phi) \equiv \frac{f_{,\phi}}
{\kappa f^{5/2} V^{3/2}}\,.
\label{vars}
\ena
From the Friedmann equation~(\ref{frieq}), we have
the constraint equation
\bea
\Omega_m \equiv \frac{\kappa^2 \rho_m}{3H^2}
=1-(1-\tg) x^2-z^2\,,
\ena
where
\bea
\tg \equiv 1/\gamma=\sqrt{1-y^2/3x^2}\,.
\ena
The energy fraction and the equation of state of the DBI
field $\phi$ are given by
\bea
&& \Omega_\phi = (1-\tg)x^2+z^2\,,\\
&& w_{\phi} = \frac{\tg(1-\tg)x^2-z^2}{(1-\tg)x^2+z^2}\,.
\ena
{}From Eq.~(\ref{frieq2}) we obtain
\bea
\frac{H'}{H}=-\frac12 y^2-\frac32 (1+w_m)
\left[ 1-(1-\tg)x^2-z^2 \right]\,,
\ena
where a prime represents a derivative with respect to
the number of e-foldings $N={\rm ln}\,a$.
The effective equation of state,
$w_{\rm eff} \equiv \frac{P_m+P_\phi}{\rho_m+\rho_\phi}= -1-2H'/3H$, is
\bea
w_{\rm eff}=-1+\frac13 y^2+
(1+w_m) \left[1-(1-\tg)x^2-z^2 \right]\,.
\ena

Taking the derivative of $x$, $y$, $z$, $\mu_1(\phi)$
and $\mu_2 (\phi)$ with respect to $N$,
we obtain the following equations:
\bea
\label{au1}
x' &=&
-\frac12 (\mu_1+\mu_2) \frac{yz^3}{x^2}
-\frac{y^2}{2x}+x \left[ \frac{y^2}{2}
+\frac32 (1+w_m) \left\{ 1-(1-\tg)x^2
-z^2 \right\} \right]\,, \\
\label{au2}
y' &=&
-\frac32 \left[ \left(1+\tg^2 \right)
\mu_1+\left( 1-\tg \right)^2 \mu_2
\right] \frac{z^3}{x}-\frac32 \left(1+\tg^2
\right)y \nn
&& + y \left[ \frac{y^2}{2}+\frac32 (1+w_m)
\left\{ 1-(1-\tg)x^2-z^2 \right\}
\right]\,, \\
\label{au3}
z' &=&
\frac12 \mu_1 \frac{yz^2}{x}+
z \left[ \frac{y^2}{2}+\frac32 (1+w_m)
\left\{ 1-(1-\tg)x^2-z^2 \right\}
\right]\,, \\
\mu_1' &=&
\mu_1^2\,\frac{yz}{x}
\left( \frac{VV_{,\phi \phi}}{{V_{,\phi}}^2}-
\frac12 \frac{V}{V_{,\phi}}
\frac{f_{,\phi}}{f}-\frac32 \right)\,,\\
\mu_2'  &=&
\mu_2^2\,\frac{yz}{x} fV
\left( \frac{ff_{,\phi \phi}}{{f_{,\phi}}^2}
-\frac32 \frac{V_{,\phi}}{V} \frac{f}{f_{,\phi}}
-\frac52 \right)\,.
\ena
If both $\mu_1$ and $\mu_2$ are constants, for example,
$V \propto f^{-1} \propto e^{\a \phi}$ where $\a$ is a constant,
the set of Eqs.~\p{au1} -- \p{au3} becomes an autonomous system.
Actually when $\mu_1$ is a constant, the potential is obtained
by integrating Eq.~\p{vars}:
\bea
\label{po}
V = \left( \frac{\kappa}{2}\mu_1
\int f^{1/2} {\rm d}\phi \right)^{-2}.
\ena
For the ADS throat ($f=\lambda/\phi^4$),
Eq.~(\ref{po}) gives
\bea
V(\phi)=\frac{4}{\kappa^2 \mu_1^2 \lambda}
\left( \frac{\phi}{1+c\phi} \right)^2\,,
\ena
where $c$ is an integration constant.
In the region $|c\phi| \ll 1$, this potential
reduces to the quadratic one:
$V(\phi) \propto \phi^2$.

In what follows, we specialize to the case of the AdS throat,
$f(\phi)=\l/\phi^4$, and the quadratic potential,
$V(\phi)=m^2\phi^2/2$. In this case, $\mu_1$ is a constant
and $\mu_2=-2\mu_1\tg x^2 z^{-2}$.
The evolution Eqs.~\p{au1} -- \p{au3}
can be written as the following autonomous system:

\bea
\label{au11}
x' &=&
-\frac12 \mu_1\left(1-2\tg\frac{x^2}{z^2}\right)
 \frac{yz^3}{x^2}
-\frac{y^2}{2x}+x \left[ \frac{y^2}{2}
+\frac32 (1+w_m) \left\{ 1-(1-\tg)x^2
-z^2 \right\} \right]\,, \\
\label{au22}
y' &=&
-\frac32 \mu_1 \left[ 1+\tg^2-2 \tg
\left( 1-\tg \right)^2 \frac{x^2}{z^2}
\right] \frac{z^3}{x}-\frac32 \left(1+\tg^2
\right)y \nn
&& + y \left[ \frac{y^2}{2}+\frac32 (1+w_m)
\left\{ 1-(1-\tg)x^2-z^2 \right\}
\right]\,, \\
\label{au33}
z' &=&
\frac12 \mu_1 \frac{yz^2}{x}+
z \left[ \frac{y^2}{2}+\frac32 (1+w_m)
\left\{ 1-(1-\tg)x^2-z^2 \right\}
\right]\,,
\ena
where $\mu_1=2\sqrt{2}/(\k \sqrt{\l}\;m)$.

%%%%%%%%%%%%%%%%%%%%%%%%%%%%%%%%%%%
\section{Scaling Solutions}
\label{sec2}
%%%%%%%%%%%%%%%%%%%%%%%%%%%%%%%%%%%

One can derive the fixed points of the system~\p{au11} -- \p{au33}
by setting $x'=0$, $y'=0$ and $z'=0$. The fixed points correspond
to an expanding universe with a scale factor $a(t)$ given by $a
\propto t^p$, where $p=2[y^2+3(1+w_m)\Omega_m]^{-1}$. From
Eq.~(\ref{au33}) we find that there are two cases: (i) $z=0$ and
(ii) $y^2+3(1+w_m)[1-(1-\tg)x^2-z^2]= -\mu_1 yz/x$. We will study
the case $\dot{\phi}<0$, i.e., $y<0$.

In the case (i) we have the following fixed points:
\begin{itemize}
\item [({\bf A})] Fluid-dominated solutions
\bea
(x,y,z)=(0,0,0),~~
\Omega_m=1,~~
w_{\rm eff}=w_m.
\ena
\item [({\bf B})] Kinetic-dominated solutions
\bea
(x,y,z)=(1,-\sqrt{3},\,0),~~
\Omega_m=0,~~
w_{\rm eff}=0.
\ena
\end{itemize}
The fixed point (A) is fluid-dominated solutions since $\Omega_m=1$.
The fixed point (B) corresponds to kinetic-dominated solutions.
They behave like dust (i.e., non-relativistic matter),
which are power-law expanding solutions with $a \propto t^{2/3}$.

In the case (ii) one has either $\mu_1(z^2-2\tg x^2)z+\mu_1 x^2z
+xy=0$ or $y=0$ from Eqs.~(\ref{au11}) and (\ref{au33}). In the
former situation, we obtain either $y^2=3x^2$ (i.e., $\tg=0$) or
$x^2(1-2\tg)=0$ by using Eq.~(\ref{au22}).
When $y^2=3x^2$, the fixed points are given by
\begin{itemize}
\item [({\bf C})] Accelerated solutions
\bea
&& x=[\mu_1 (\sqrt{\mu_1^2+12}-\mu_1)/6]^{1/2}\,,\nn
&& y=-\sqrt{3}x\,,\nn
&& z=\sqrt{3}(\sqrt{\mu_1^2+12}-\mu_1)/6\,,\nn
&& \Omega_m=0,~~
w_{\rm eff}
=-1+\mu_1 (\sqrt{\mu_1^2+12}-\mu_1)/6.
\ena
\item [({\bf D})] Scaling solutions
\bea
&& x=[-3(1+w_m)^3/(w_m \mu_1^2)]^{1/2}\,,\nn
&& y=-\sqrt{3}x\,,\nn
&& z=\sqrt{3}(1+w_m)/\mu_1,\nn
&& \Omega_m=1+3(1+w_m)^2/(w_m \mu_1^2),~~
w_{\rm eff}=w_m.
\ena
\end{itemize}
Both the fixed points (C) and (D) exist in the ultra-relativistic
region: $\gamma \to \infty$.
These solutions are chosen by the condition that $x>0$,
$z>0$ and $\Omega_m \geq 0$ in the expanding universe.
This requires $-1<w_m<0$ and $\mu_1>\sqrt{-3/w_m}\,(1+w_m)$ in (D).
The fixed point (C) leads to an accelerated expansion for $\mu_1<2$,
which was proposed as an alternative to the slow-roll
inflation~\cite{sil03,che05}.
In such models inflation may also proceed when the field is
rolling relatively fast.
The fixed point (D) corresponds to scaling solutions in which the
ratio of their densities is a non-trivial constant.
Note that even when $\mu_1$ changes with time
the fixed points (C) and (D) can be regarded as ``instantaneous'' fixed points.

Under the condition $\tg=0$, the relation
$x^2(1-2\tg)=0$ gives a fixed point which
is not much different from the point (A).
Since an accelerated expansion is not realized, this case
is out of our interest.

\begin{figure}[t]
\begin{center}
\includegraphics[width=12cm]{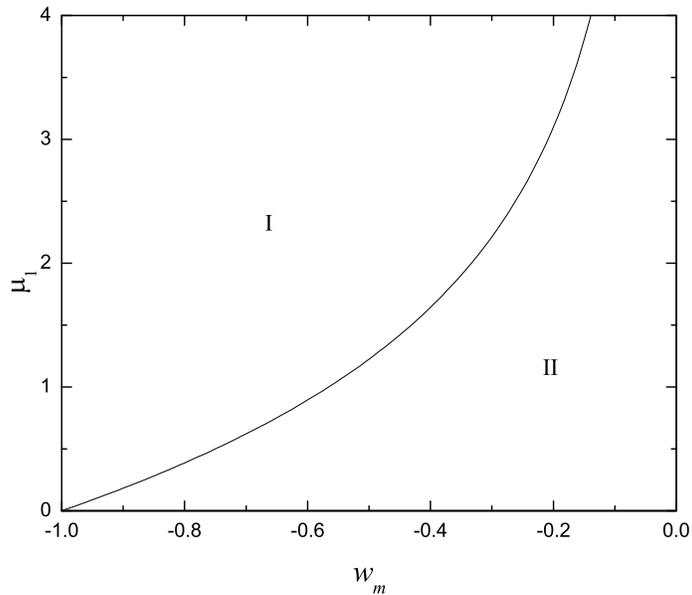}
\end{center}
\caption{Stable regions in the $(w_m, \mu_1)$ parameter space.
In the region I, all fixed points exist and the fixed point (D)
is the attractor solution. In the region II, the fixed point (D)
does not exist and the fixed point (C) is the
attractor solution.}
\label{fig}
\end{figure}

In order to analyze their stability, we substitute
linear perturbations about the fixed points into the
field equations~\p{au11} -- \p{au33}. To the first
order in the perturbations, we obtain two independent
equations of motion for $\tg=0$.
If their eigenvalues are both negative, the fixed point is stable.
For the fixed point (A), we get two eigenvalues
\bea
\l_1=3(1+w_m)/2\,, \quad \l_2=3w_m,
\ena
which indicate that it is unstable if $-1<w_m<1$.
For the fixed point (B), we get two eigenvalues
\bea
\l_1=3/2\,, \quad \l_2=-3w_m/2,
\ena
which indicate that it is also unstable.
For the fixed point (C), two eigenvalues are
\bea
&& \l_1= -\frac14\sqrt{\mu_1^2+12}\left(\sqrt{\mu_1^2+12}-\mu_1\right),\nn
&& \l_2= -\frac14\left[6(1+w_m)+\mu_1^2-\mu_1\sqrt{\mu_1^2+12}\,\right],
\ena
which indicate that it is stable for $\mu_1<\sqrt{-3/w_m}\,(1+w_m)$.
For the point (D), two eigenvalues are
\bea
&& \l_1= -\frac34\left[1-w_m+\sqrt{24(1+w_m)^3/\mu_1^2+ (3w_m+1)^2}\,\right],\nn
&& \l_2= -\frac34\left[1-w_m-\sqrt{24(1+w_m)^3/\mu_1^2+ (3w_m+1)^2}\,\right].
\ena
Thus the scaling solutions are always stable when they exist
for $\mu_1>\sqrt{-3/w_m}\,(1+w_m)$.
The different regions in the $(w_m, \mu_1)$ parameter space
lead to different qualitative evolution in Fig.~\ref{fig}.
In the region I, all four fixed points exist and the fixed point (D)
is the attractor solution. In the region II, the fixed point (D)
does not exist and the fixed point (C) is the
attractor solution.

%%%%%%%%%%%%%%%%%%%%%%%%%%%%%%%%%%%
\section{Conclusions and Discussions}
\label{sec3}
%%%%%%%%%%%%%%%%%%%%%%%%%%%%%%%%%%%

We have investigated the cosmological evolution for a
spatially-flat FRW universe containing a Dirac-Born-Infeld
field and a perfect fluid.
We find that the field equations can be cast into an autonomous
system~\p{au11} -- \p{au33} in the case of the AdS throat and
the quadratic potential.
In addition to the DBI inflationary solutions (C),
there exist scaling solutions (D) in which the ratio of the
energy densities of the two components is a constant.
We have analyzed the existence and stability of the fixed points,
and shown that the scaling solutions (D) exist and are stable
when the equation of state of the perfect fluid satisfies
$-1<w_m<0$, for $\mu_1>\sqrt{-3/w_m}\,(1+w_m)$ located in the
region I of the parameter space, and in the ultra-relativistic
regime (i.e., $\tg=0$).

There is another string-motivated choice of the warp
factor, i.e., a constant $f$. This corresponds to the case in
which inflation proceeds in the angular directions instead
of in the radial~\cite{paj08}. In this case $\mu_2$ vanishes
and $\mu_1$ becomes a constant for an inverse square potential.
In the ultra-relativistic region all results are the same as
those derived above.
Given a warp factor $f(\phi)$ and a potential term $V(\phi)$,
in principle, the set of equations~\p{au1} -- \p{au3} can be
written as an autonomous system since both $\mu_1(\phi)$ and
$\mu_2(\phi)$ in the equation set can be expressed in terms
of the variables $x$ and $z$. It is worth studying further
cosmological dynamics of general functions $f(\phi)$ and
$V(\phi)$ to explain for the present acceleration of the universe.
We mention that the dynamics of tachyon actions with a runaway
potential contain caustics with multi-valued regions because high
order spatial derivatives of the tachyon field become
divergent~\cite{fel02}. Here we do not consider a runaway potential
but a quadratic one, which may stabilize the system~(\ref{action}).
To check if this expectation is really true or not is an interesting
problem, which we leave for future study.

%%%%%%%%%%%%%%%%%%%%%%%%%%%%%%%%
%%%%%
\section*{Acknowledgements}
We would like to thank Shinji Tsujikawa and Sudhakar Panda
for useful discussions.
This work was supported in part by the Grant-in-Aid for
Scientific Research Fund of the JSPS Nos. 16540250 and 06042.
%%%%%%%%%%%%%%%%%%%%%%%%%%%%%%%%%%%%%

\end{document}